\documentclass[prl,twocolumn,showpacs,superscriptaddress,amsmath,amssymb]{revtex4}
\usepackage{graphicx}% Include figure files
\usepackage{bm}% bold math
\usepackage{amsmath} % AMSmath for align environment etc.

\def\up{\text{\mbox{$\uparrow$}}}       % This makes the up and down arrows look better
\def\down{\text{\mbox{$\downarrow$}}}   %

\begin{document}

\title{Two spin measurements in exchange interaction quantum computers}

\author{S. D. Barrett}
\email{sean.barrett@hp.com}
\affiliation{Hewlett-Packard
Laboratories, Filton Road, Stoke Gifford, Bristol BS34 8QZ, U.K.}
\affiliation{Semiconductor Physics Group, Cavendish Laboratory,
University of Cambridge, Madingley Road, Cambridge CB3 0HE, U.K.}

\author{T. M. Stace}
\affiliation{Semiconductor Physics Group, Cavendish Laboratory,
University of Cambridge, Madingley Road, Cambridge CB3 0HE, U.K.}

\date{\today}
% 03.65.Ta Foundations of quantum mechanics; measurement theory
% 03.67.Lx -Quantum computation
% 73.21.La Quantum dots (Electron states and collective excitations in multilayers, quantum wells, mesoscopic, and nanoscale systems)
% 72.25.Rb Spin relaxation and scattering

\begin{abstract}
We propose and analyse a method for single shot measurement of the
total spin of a two electron system in a coupled quantum dot or
donor impurity structure, which can be used for readout in a
quantum computer. The spin can be inferred by observing spin
dependent fluctuations of charge between the two sites, via a
nearby electrometer. Realistic experimental parameters indicate
that the fidelity of the measurement can be larger than 0.9999
with existing or near-future technology.
\end{abstract}
\maketitle

Semiconductor technology is rapidly reaching the level where
quantum systems comprising of one or two electron spins can be
confined and coherently manipulated within a single nanostructure
\citep{Fujisawa2002,Elzerman2004}. These experiments are important
from a fundamental point of view, and will also pave the way to
new applications in quantum information processing (QIP)
\citep{Loss98,Kane1998,Vrijen2000}. A key requirement for these
QIP schemes is the realization of a single shot readout technique,
whereby the spin state of a one- or two-electron system can be
determined in a single measurement run. Such measurements are
important both for the ongoing experimental development of spin
qubit systems (e.g. for characterizing qubit parameters, and
studying the physics of decoherence processes) and, ultimately,
for developing scalable QIP architectures.

% Other schemes. Segue into problems with existing schemes
% for all electrical spin readout, esp. in donor impurity schemes.

Recently, elegant experiments have demonstrated single shot
readout of single electron spins, via optical techniques in a
defect center in diamond \cite{Jelezko2004}, and using an
all-electrical technique involving spin to charge conversion in a
quantum dot, together with charge detection with a nearby quantum
point contact (QPC) \citep{Elzerman2004}. A number of other
methods based on spin-to-charge conversion have also been proposed
\cite{Loss98,Kane1998,Engel2004,OtherSpinReadoutSchemes}.
Of particular interest are readout schemes in donor impurity
implementations of QIP \cite{Kane1998,Vrijen2000}, in which an
electric field is applied to induce spin-dependent polarization of
a two-electron double-donor system. It is believed that these
proposals may fail due to the short lifetime of the quasi-bound
two electron system under the large $E$ field required to observe
a significant polarization \cite{SeansThesis,Hollenberg2004}.
In this paper we discuss an alternative scheme, whereby the total
spin of a two electron system can be inferred by observing spin
dependent \emph{fluctuations} of charge between two tunnel coupled
quantum dots (CQD) \footnote{Note that our scheme is applicable to
both quantum dot or donor impurity based implementations, although
hereafter we will refer to the electron sites as `dots'.}, even in
the absence of an external electric field. Our scheme therefore
avoids the lifetime issue in donor impurity systems, and
furthermore may be easily integrated into proposed QIP
architectures, as it does not require magnetic tunnel barriers,
tunnel coupling to electron reservoirs, or external R.F. fields.
The scheme can be used for read out of a single spin qubit
(together with an ancilla spin), read out of an encoded qubit
\cite{DiVincenzo2000}, or, together with appropriate single spin
operations, as a two-qubit readout device.

\begin{figure}
\includegraphics[width=0.3\textwidth]{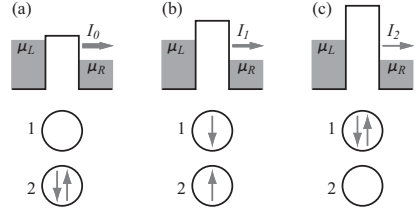}
\caption{A schematic of the singlet-triplet readout scheme.
Configurations (a) and (c) are energetically forbidden for triplet
states, and thus monitoring fluctuations in the QPC current allows
a measurement in the singlet-triplet basis.} \label{fig:Device}
\end{figure}

Spin dependent charge fluctuations can be observed by continuously
measuring the CQD system with an electrometer adjacent to one of
the dots (see Fig. \ref{fig:Device}). For the purposes of this
work, we model the electrometer as a QPC, although it might
equally be a single electron transistor.  Note that the QPC has a
dual role in this scheme: it acts both as a noise source which can
induce inelastic transitions in the CQD system, and as a detector
to observe these transitions.

%5
In what follows, we first describe a model for the coupled CQD-QPC
system, and subsequently derive master equations for both the
unconditional and conditional dynamics of the CQD system. We use
these results to simulate individual runs of the measurement
scheme, and to characterize the detector output for the different
measurement outcomes. We show that observing the detector output
leads to a quantum measurement in the singlet-triplet basis: for a
singlet state (total spin $S=0$) one observes fluctuations in the
output current, while for a triplet state ($S=1$), these
fluctuations are energetically suppressed, owing to the Pauli
principle. Finally we determine the single-shot singlet-triplet
measurement time and fidelity for realistic experimental
parameters.

%6
\emph{Model.} A number of authors have considered the problem of a
continuously observed \emph{single} charge in a coupled dot system
\cite{OtherPapersOnChargeQubitMeasurement,Goan2001b,Engel2004,StaceBarrett2003a,StaceBarrett2003b}.
Here, we adopt the quantum trajectories approach of
\cite{StaceBarrett2003a,StaceBarrett2003b}, which takes into
account the role of energy exchange mechanisms between the
observed system and a QPC detector at finite voltage bias. We
describe the two-electron coupled-dot system by a two site Hubbard
Hamiltonian of the form
\begin{equation}
H = \hbar \Delta(t)\sum_{\sigma =\up, \down} (a_{1
\sigma}^{\dagger}a_{2 \sigma}^{\phantom{\dagger}} + a_{2
\sigma}^{\dagger}a_{1 \sigma}^{\phantom{\dagger}} ) + \hbar
U\sum_{i = 1, 2}n_{i\down}n_{i\up} \, ,
\label{eq:HubbardHamiltonian}
\end{equation}
where $a^{(\dagger)}_{i\sigma}$ is the fermionic annihilation
(creation) operator for an electron on site $i$ with spin
$\sigma$, $n_{i\sigma} = a^{\dagger}_{i\sigma}a_{i\sigma}$,
$\Delta(t)$ is the tunnelling amplitude between sites, and $U$ is
the on-site Coulomb energy. We allow $\Delta(t)$ to be time
dependent, since this can lead to shorter measurement times (see
below), although in what follows, we frequently suppress this time
dependence for clarity. $H$ is spanned by the four singly occupied
states $\vert \sigma_1 \sigma_2\rangle$ (where $\sigma_i =
\up,\down$ denotes the spin on dot $i$) and two doubly occupied
states $\vert d_i \rangle = a^{\dagger}_{i\up}a^{\dagger}_{i\down}
\vert 0 \rangle$ ($i = 1,2$). The eigenstates of $H$ are as
follows. The singly occupied triplet states $\vert t_+ \rangle =
\vert \up \up \rangle$, $\vert t_0 \rangle = 2^{-\frac{1}{2}}
(\vert \up \down \rangle + \vert \down \up \rangle)$ and $\vert
t_- \rangle = \vert \down \down \rangle$ form a degenerate
subspace with eigenvalue $0$. The remaining states (with total
spin $S=0$) are $\vert s_0 \rangle
 =2^{-\frac{1}{2}} \cos(\theta/2) (\vert
\up \down \rangle - \vert \down \up \rangle) + 2^{-\frac{1}{2}}
\sin(\theta/2) (\vert d_1 \rangle + \vert d_2 \rangle)$, $\vert
s_1 \rangle
 =2^{-\frac{1}{2}} \sin(\theta/2) (\vert
\up \down \rangle - \vert \down \up \rangle) + 2^{-\frac{1}{2}}
\cos(\theta/2) (\vert d_1 \rangle + \vert d_2 \rangle)$, and
$\vert s_2 \rangle = 2^{-\frac{1}{2}}(\vert d_1 \rangle - \vert
d_2 \rangle)$, where $\theta = \tan^{-1} (4 \Delta/U)$. These
states have eigenvalues $-J$, $U+J$ and $U$, respectively, where
$J(t) = (\sqrt{U^2 + 16 \Delta^2(t)} - U)/2 \approx 4
\Delta^2(t)/U$ is the exchange splitting between the lowest energy
singlet and triplet states. $J(t)$ can be varied over many orders
of magnitude by varying a surface gate voltage
\citep{Loss98,Wellard2003}.

The entire system (i.e. CQD coupled to a QPC) has a Hamiltonian
$H_{\mathrm{tot}} = H + H_{\mathrm{tun}} + H_{\mathrm{leads}}$
where
\begin{align}
H_{\mathrm{tun}}   & =   \hbar \sum_{k,q,\sigma} \left(T_{kq} +
\chi_{kq} n_1 \right)
a^\dag_{Lk\sigma}a^{\phantom{\dag}}_{Rq\sigma} + \mathrm{h.c.} \,, \label{eq:TunnellingHamiltonian} \\
H_{\mathrm{leads}} & = \hbar \sum_{k,\sigma} \omega_{Lk}
a^\dag_{Lk\sigma}a^{\phantom{\dag}}_{Lk\sigma} + \hbar
\sum_{q,\sigma}
\omega_{Rq}a^\dag_{Rq\sigma}a^{\phantom{\dag}}_{Rq\sigma} \,.
\label{eq:ReservoirHamiltonian}
\end{align}
Here, $a^{\phantom{\dag}}_{Lk\sigma}$
($a^{\phantom{\dag}}_{Rq\sigma}$) are fermionic annihilation
operators for electrons in the $k$th ($q$th) mode on the left
(right) side of the barrier with spin $\sigma$ and angular
frequency $\omega_{Lk}$ ($\omega_{Rq}$). Tunnelling between the
left and right baths is described by the factor $T_{kq} +
\chi_{kq} n_1$, where $n_1 = n_{1\up} + n_{1\down}$ is the total
occupancy of dot `1'. $T_{kq} \approx T_{\mathrm{av}}$ and
$\chi_{kq} \approx \chi_{\mathrm{av}}$ are assumed to vary slowly
over the energy range where tunnelling is allowed.

\emph{Unconditional master equation.} We now derive an
unconditional master equation (UME) for the evolution of the
reduced density matrix of the CQD system, $\rho(t)$. The UME is
obtained by first transforming to an interaction picture in which
the dynamics are governed by $H_I(t) = U^{\dag}(t)
H_{\mathrm{tun}} U(t)$, with $U(t) = e^{-i(H + H_{\mathrm{leads}})
t}$. In the Born and Markov approximations \cite{Breuer2002}, the
resulting dynamics for the CQD system is given by
\begin{equation}
\dot{\rho}_I = \mathrm{tr}_{\mathrm{leads}} \{ - \frac{1}{\hbar^2}    %
\int_{-\infty}^t dt' [H_I(t),[H_I(t'),\rho_I(t) \otimes \rho_{L}
\otimes \rho_{R}]]\}. \label{eq:BornMarkov}
\end{equation}
Here, $\rho_I(t) = U^{\dag}(t) \rho(t) U(t)$, and $\rho_{L}$ and
$\rho_{R}$ are the lead density matrices, given by Fermi-Dirac
distributions with chemical potentials $\mu_L$ and $\mu_R$. The
interaction picture tunnelling Hamiltonian may be written
explicitly as $H_I(t) = \sum_{k,q,\sigma}[T_{kq}+\chi_{kq} n_1(t)]
e^{i(\omega_{Lk\sigma} - \omega_{Rq\sigma})t}
a^{\dag}_{Lk\sigma}a_{Rq\sigma}+\mathrm{h.c.}$, where the time
dependence of $n_1(t)$ is given by $n_1(t) = \frac{1}{2} -
e^{-i(J+U)t} \sin (\theta/2)  \vert s_0 \rangle\langle s_2 \vert -
e^{i J t} \cos (\theta/2) \vert s_1 \rangle\langle s_2 \vert
+\mathrm{h.c.}$.  Note that Eq. (\ref{eq:BornMarkov}) corresponds
exactly to the first non-vanishing term in a perturbative
expansion of $\rho_I(t)$ in terms of $H_I(t)$ \cite{Breuer2002}.

We now make two further controlled approximations. First, we make
a rotating wave approximation, setting very rapidly oscillating
factors $e^{i\alpha_i t} \to 0$, where $\alpha_i = \{ U,J+U \}$.
This is valid for $\alpha_i \gg \nu^2 V$, where $\nu=\sqrt{4 \pi
g_L g_R}\chi_{\mathrm{av}}$ is a dimensionless coupling constant,
$g_i$ is the density of states in lead $i$, and $V =
(\mu_L-\mu_R)/\hbar$. Second, we neglect small terms of order
$\nu^2 \sqrt{JV}$, which is valid when  $J \ll V$. At low
temperatures ($k_B T \ll \hbar V$), and taking $V \ge J+U$, the
UME is
\begin{equation}
\dot{\rho}(t)  =  -\frac{i}{\hbar} \left[H , \rho(t)\right] +
\sum_{i=1}^3 \mathcal{D}[c_i] \rho(t) \equiv \mathcal{L}
\rho(t)\,, \label{eq:MasterEquation}
\end{equation}
where $\mathcal{D}[c]\rho = \mathcal{J}[c]\rho -
\mathcal{A}[c]\rho$, $\mathcal{J}[c]\rho = c\rho c^\dag$,
$\mathcal{A}[c]\rho = (c^\dag c \rho + \rho c^\dag c)/2$. The
Lindblad operators are given by
\begin{align}
c_1   & = \nu \sqrt{V - (J+U)}  \sin
\frac{\theta}{2} \vert s_2 \rangle\langle s_0 \vert   \, , \label{eq:C1} \\
c_2   & = \nu \sqrt{V + (J+U)}  \sin \frac{\theta}{2}
\vert s_0 \rangle\langle s_2 \vert \, , \label{eq:C2} \\
c_3   & = \sqrt{V} \left[\mathcal{T} + \nu - \nu\cos
\frac{\theta}{2} (\vert s_1 \rangle\langle s_2 \vert +
\mathrm{h.c.}) \right] \,, \label{eq:C3}
\end{align}
where  $\mathcal{T} =\sqrt{4 \pi g_L g_R} T_{\mathrm{av}}$. The
$c_i$'s can be associated with different types of tunnelling
process in the QPC. $c_{1,2}$ correspond to inelastic transitions,
in which electrons tunnelling through the QPC exchange energy with
the CQD system, and are accompanied by transitions between the
low-energy, singly occupied state, and the high energy, doubly
occupied state. $c_{3}$ corresponds to a quasi-elastic transition,
in which electrons tunnel through the QPC without changing energy.

\emph{Conditional dynamics.} Individual measurement runs can be
simulated using a conditional master equation (CME), which
describes the evolution of the CQD system conditioned by the
observed detector output, and also permits calculation of the
current power spectrum. To derive the CME, we use an explicit
model of the measurement process in terms of projective
measurements of the number of electrons that have tunnelled
through the PC, similar to that presented in
\cite{StaceBarrett2003b,Breuer2002}. The CME is found to be
\begin{multline}
d{\rho_c}(t) = dN_c(t) \left[
\frac{\sum_{n=1}^{3}\mathcal{J}[c_n]}{P_{1c}(t)} - 1 \right] \rho_c(t) \\
+  dt \left\{ -\sum_n\mathcal{A}[c_n] \rho_c(t) + P_{1c}(t) + i
[H_{\mathrm{qb}}, \rho_c(t)] \right\}. \label{eq:CME}
\end{multline}
Here, $dN_c(t)$ is a stochastic point process, corresponding to
the number of electrons (0 or 1) which tunnel through the QPC in
an interval $dt$, $P_{1c}(t)
dt\equiv\mathrm{tr}\left\{\sum_n\mathcal{J}[c_n]\rho_c(t)\right\}
dt$ is the corresponding probability of observing such a
tunnelling event. Note that, in deriving Eqs.
(\ref{eq:MasterEquation}) and (\ref{eq:CME}), we have used a
course graining in time, which means that these equations are
valid for timescales longer than $U^{-1}$. In practice $U \gtrsim
10^{12}$ s$^{-1}$, and so this is not a significant restriction.

\begin{figure}
\includegraphics{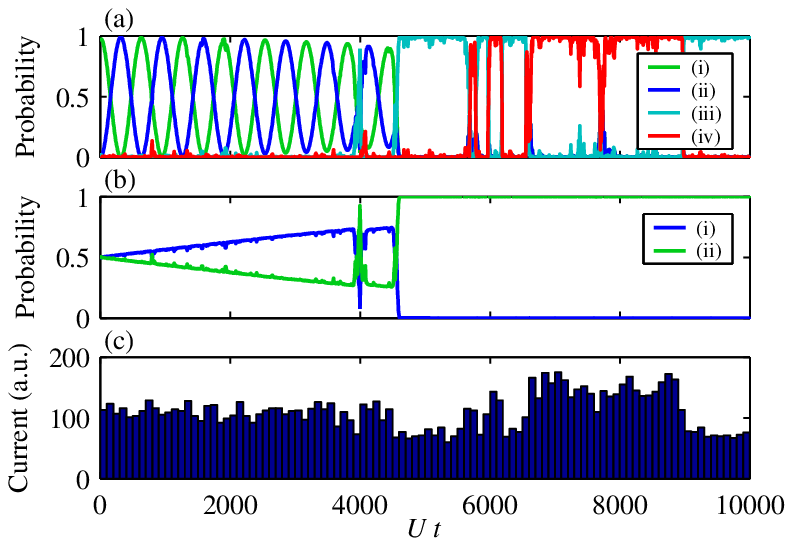}
\caption{
            A sample evolution of the conditional density matrix,
            $\rho_c$, during a single measurement run. %%
            (a) Curves $i=$ \{(i) \ldots (iv)\} denote the
            probabilities $P(i,t) = \mathrm{tr}[\vert \phi_i
            \rangle\langle \phi_i \vert \rho_c(t)]$ for the system
            to be in the state $\vert \phi_i \rangle\langle \phi_i
            \vert$ at time $t$, where $\vert \phi_i \rangle =
            \left \{\vert \up \down\rangle,\,\, \vert
            \down \up \rangle,\,\,\vert d_1
            \rangle,\,\,\vert d_2 \rangle\right\}$.
            (b) Corresponding probabilities for the system to be in a
            triplet state (curve (i)) or singlet state (curve (ii)).
            (c) Corresponding QPC current, as a histogram of tunnelling events.
            The parameters used in this simulation are $\Delta = 0.05
            U$ (we keep $\Delta(t)$ constant for this simulation), $V = 5 U$, $T = \sqrt{0.3} \approx 0.54$, $\nu = -0.15 T \approx
            -0.082$. The initial state is taken to be
            $\rho_c = \vert \up \down \rangle\langle \up \down \vert$. In this
            measurement run, the final state of the system is a singlet state.
            }
            \label{fig:SampleEvolSinglet}
%
% t=0.05;
% U=1.0; J= 0.5*(sqrt(16*t^2+U^2)-U);
%
% T = 0.5477, U = 1, V = 5, nu = -0.0822
%
% i.e.:
% V = 5*U;
% T = sqrt(0.3); nu = -0.15*T;
\end{figure}

\begin{figure}
\includegraphics{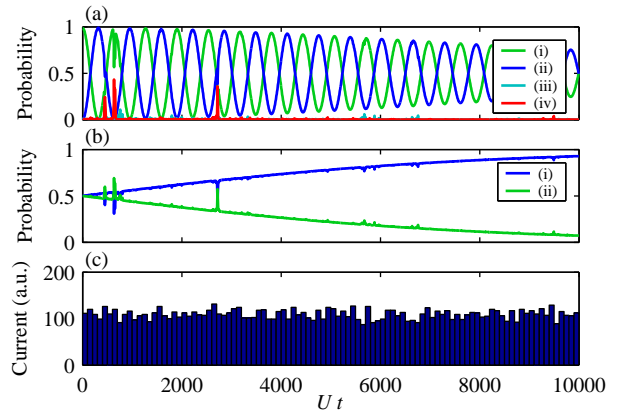}
\caption{
            Another sample evolution. In this measurement run, the state of the
            system becomes localized in the triplet subspace.
            Parameters as
            in Fig. \ref{fig:SampleEvolSinglet}.
            }
 \label{fig:SampleEvolTriplet}
\end{figure}

%8
Two sample solutions of Eq. (\ref{eq:CME}) are shown in Figs.
\ref{fig:SampleEvolSinglet} and \ref{fig:SampleEvolTriplet}. In
Fig. \ref{fig:SampleEvolSinglet}(a) the system initially undergoes
almost coherent oscillations between the singly occupied states
$\vert \up \down \rangle\langle \up \down \vert$ and $\vert \down
\up \rangle\langle \down \up \vert$ at the exchange frequency $J$.
After a certain time the dynamics undergoes a qualitative change:
the system suddenly jumps into the doubly occupied state $\vert
d_1 \rangle\langle d_1 \vert$, and thereafter undergoes stochastic
jumps between states with distinct charge configurations. This
behaviour is reflected in the QPC output [Fig.
\ref{fig:SampleEvolSinglet}(c)]: when both electrons are localized
on dot `1' (`2'), the current is smaller (larger) than when the
system is in a singly occupied state. This is reminiscent of a
random telegraph switching (RTS) signal. At the first jump, the
system becomes localized in the singlet subspace [Fig.
\ref{fig:SampleEvolSinglet}(b)]. This is because the doubly
occupied state $\vert d_1 \rangle\langle d_1 \vert$ is a
spin-singlet state. Doubly occupied triplet states are
energetically forbidden, and therefore when a transition into the
doubly occupied subspace is observed, one can infer that the total
spin of the system is a spin-singlet. In Fig.
\ref{fig:SampleEvolTriplet}, no jump into the doubly occupied
subspace is observed, and the detector output current remains
constant (up to shot noise). This leads to a gradual increase in
the observer's confidence that the CQD system is in a spin-triplet
state [Fig. \ref{fig:SampleEvolTriplet}(b)]. In both cases, the
final state of the system is localized in either the singlet or
triplet subspace, and a strong measurement of the total spin of
the system in the singlet-triplet basis is obtained.

\emph{Detector power spectra.} The possible detector outputs can
be characterized by the power spectrum of the current, $I(t)$,
through the QPC. The power spectrum is given by $S(\omega)=2
\int_{-\infty}^{\infty} d\tau G(\tau) e^{-i\omega t}$, where
$G(\tau) = E[I(t+\tau)I(t)]-E[I(t+\tau)]E[I(t)]$, and $E[\ldots]$
denotes the classical expectation. Following Refs.
\cite{Goan2001b,StaceBarrett2003a,StaceBarrett2003b}, we have
$G(\tau) = e^2 [\mathrm{tr}\{ \sum_{n,n'}
\mathcal{J}[c_n]e^{\mathcal{L}\tau}\mathcal{J}[c_n']\rho_{\infty}
\} -
\mathrm{tr}\left\{\sum_{n}\mathcal{J}[c_n]\rho_{\infty}\right\}^2
+
\mathrm{tr}\left\{\sum_{n}\mathcal{J}[c_n]\rho_{\infty}\right\}\delta(\tau)]
$, where $\rho_{\infty}$ is a steady state solution of the UME.
Spectra corresponding to the different measurement outcomes can be
found by evaluating $G(\tau)$ for different $\rho_{\infty}$,
corresponding to steady states localized in the singlet or triplet
subspaces.

\begin{figure}
\includegraphics{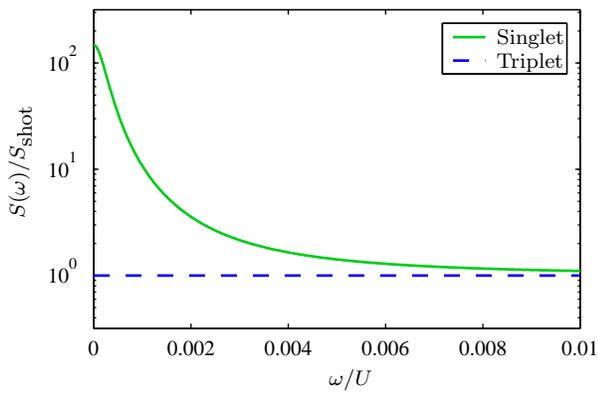}
\caption{Normalized steady state power spectra for the QPC
detector current for both singlet and triplet steady states.
Parameters as in Fig. \ref{fig:SampleEvolSinglet}.}
\label{fig:PowerSpectrum}
\end{figure}

For a singlet state outcome, the steady state is $\rho_{S,\infty}
= [(V+U+J)\vert s_0 \rangle\langle s_0 \vert + (V-U-J)\vert s_1
\rangle\langle s_1 \vert + (V-U-J)\vert s_2 \rangle\langle s_2
\vert]/(3V-U-J)$. In the parameter regime of interest ($\omega, J
, \nu^2 V \ll J+U \le V $), the power spectrum is well
approximated by
\begin{equation}
S_S(\omega)  = 2e\bar{I}+ \frac{16 \delta I^2  J^2 \nu^2V
\left(\frac{V-U-J}{3V-U-J}\right)}{(\omega^2 - J^2)^2 + 4\omega^2
(\nu^2 V)^2 } ,\label{eq:SingletPowerSpectrum}
\end{equation}
where $\bar{I} = e^2(\mathcal{T}+\nu)^2V$ is the average current
through the detector, and $\delta I = e^2(2 \mathcal{T} \nu)V$. In
the limit $J \ll \nu^2 V$, Eq.(\ref{eq:SingletPowerSpectrum})
corresponds to a RTS process with switching rate
$t_{\mathrm{RTS}}^{-1} = J^2/(4 \nu^2 V)$ which is the rate at
which the system jumps between the states $\vert d_1 \rangle$ and
$\vert d_2 \rangle$. For a triplet state outcome, the steady state
is an arbitrary state in the (singly occupied) triplet subspace.
The power spectrum only contains a shot noise component, and is
given by $S_T(\omega)  = 2e\bar{I}$. Power spectra for both
singlet and triplet outcomes are illustrated in Fig.
\ref{fig:PowerSpectrum}.

\emph{Measurement time and fidelity.} The total time required to
perform a single run of the singlet-triplet measurement scheme,
$t_{\mathrm{meas}}$ has two components, $t_{\mathrm{meas}} =
t_{\mathrm{rel}} + t_{\mathrm{det}}$, where $t_{\mathrm{rel}}$ is
the time taken for a singlet state to relax into the doubly
occupied subspace, and $t_{\mathrm{det}}= t_{\mathrm{RTS}} \approx
4 \nu^2 V/J^2$ is the time required to determine the presence or
absence of the RTS signal. By analysing Eq.
(\ref{eq:MasterEquation}), we find $t_{\mathrm{rel}}^{-1} \approx
\nu^2 J (V-U)/U$. $t_{\mathrm{meas}}$ can be minimized by varying
$J(t)$ over the course of the measurement, such that different
values of $J(t)$ are obtained during the `relaxation' and
`detection' phases of the measurement. In order to minimize the
$t_{\mathrm{rel}}$, $J_{\mathrm{rel}}$ should be as large as
possible, but to obtain a strong RTS signal, we require
$J_{\mathrm{det}} \lesssim \nu^2 V$. We now evaluate
$t_{\mathrm{rel}}$ and $t_{\mathrm{det}}$ for two possible
implementations \citep{Kane1998,Loss98}. For P donor qubits in Si
\citep{Kane1998}, we take $U = 43.8$ meV, $V = 45.4$ meV
\footnote{In order to prevent direct ionization of the donors by
the detector noise, we require $V \le E_{\mathrm{bind}}$, where
$E_{\mathrm{bind}} = 45.5$ meV is the single electron binding
energy for phosphorous donors in silicon.},
$\nu^2 = 6 \times 10^{-5}$ (corresponding to a change in
conductance of 5 \% for each electron added to dot `1', assuming
$\mathcal{T} = 0.3$), $J_{\mathrm{rel}} = 1.0$ meV and
$J_{\mathrm{det}} = 0.8\, \nu^2 V = 2.2$ $\mu$eV
\citep{Wellard2003}, which gives $t_{\mathrm{rel}} = 300$ ns and
$t_{\mathrm{det}} = 1.51$ ns.  For quantum dot qubits in GaAs
\citep{Loss98}, we take $U = 1$ meV, $V = 2$ meV, $\nu^2 = 6
\times 10^{-5}$, $J_{\mathrm{rel}} = 0.1$ meV and
$J_{\mathrm{det}} = 0.8\, \nu^2 V = 0.11$ $\mu$eV, giving
$t_{\mathrm{rel}} = 110$ ns and $t_{\mathrm{det}} = 34.4$ ns.

The fidelity of the measurement is approximately $F \sim 1 -
t_{\mathrm{meas}} / t_{\mathrm{mix}}$, where $t_{\mathrm{mix}}$ is
the \emph{mixing time} for unwanted transitions between the
singlet and triplet subspaces. Owing to the form of the
detector-dot interaction in Eq.(\ref{eq:TunnellingHamiltonian}),
the detector back-action does not induce such mixing transitions,
and therefore the dominant contribution to $t_{\mathrm{mix}}$ is
due to interactions with the environment. Although a lower bound
of 200 $\mu$s for the singlet-triplet relaxation time in a
\emph{single} quantum dot has been measured \citep{Fujisawa2002},
presently no data exists for the two electron-two dot case.
Therefore we take $t_{\mathrm{mix}} \sim \mathrm{min}[T_1,T_2]$,
where $T_1$ and $T_2$ are the single spin relaxation and dephasing
times, respectively (both processes can lead to transitions
between the singlet and triplet subspaces). For P donors in Si,
$T_1$ can be hours \citep{Feher1950b}, and $T_2 = 50$ ms was
recently measured \citep{Tyryshkin2003}, giving $F \sim 1 - 5
\times 10^{-6}$. For GaAs qubits, $T_1 = 0.8$ ms has been measured
\citep{Elzerman2004}, and it is believed that $T_2 \approx T_1$
\citep{Golovach2004}; thus $F \sim 1 - 10^{-4}$. These results are
particularly promising for the implementation of single shot
singlet-triplet measurements with the fidelity required for fault
tolerant quantum computation \citep{Steane2003}.

We thank  C. H. W. Barnes, W. J. Munro, T. P. Spiller and G. J.
Milburn for useful conversations. SDB thanks the EPSRC and the
E.U. NANOMAGICQ project (Contract no. IST-2001-33186) for
financial support. TMS thanks the Hackett Scholarships Committee
and CVCP for financial support.

%
%------------------------------------------------------------------------
% REFERENCES
%------------------------------------------------------------------------
%
\bibliography{srpaperetalFolded}

\end{document}